\documentstyle[12pt]{article}
\newcommand{\al}{\alpha}

\newcommand{\be}{\begin{equation}}
\newcommand{\ee}{\end{equation}}
\newcommand{\ba}{\begin{eqnarray*}}
\newcommand{\ea}{\end{eqnarray*}}

\title{Generalization of Agranovich-Toshich transformation and
constraint free bosonic representation for systems
of truncated oscillators.
}
\author{A.V.Ilinskaia
\thanks{On leave of absence from
Institute of Physics, St-Petersburg University,
191904 St-Petersburg, Russia
}
$\quad$ and $\quad$
K.N.Ilinski
\thanks{On leave of absence from
School of Physics and Space Research,
University of Birmingham, Birmingham B15 2TT,
UK; e-mail:kni@th.ph.bham.ac.uk} \\
{\small\it International Center for Theoretical Physics,}
\\
{\small\it PO Box 586, Miramare Strada Costiera 11,} \\
	    {\small\it 341000 Trieste, Italy}}
\date{  }

\begin{document}
\setcounter{page}{0}
\maketitle
\vskip -9.5cm
\vskip 9.5cm
\thispagestyle{empty}
\begin{abstract}
The generalization of
Agranovich-Toshich representation of paulion operators in terms of bosonic
ones for the case of truncated oscillators of higher ranks is
represented. We use this generalization to introduce a new constraint free
bosonic description of truncated oscillator systems. The corresponding
functional integral representations for thermodynamic quantities are given
and the application to investigations of Long Rang Order in the
system is discussed.
\end{abstract}

\vspace{1cm}

\newpage

\section{Introduction}
About three decades ago in the paper~\cite{AT} Agranovich and Toshich
proposed the bosonic expression for the creation and annihilation
operators of paulions (i.e. particles obeyed the fermionic
anticommutation relations on same sites and the bosonic commutation
relations on
different sites; alternatively, it is possible to realize them as
lattice spins $1/2$ or truncated oscillators of rank~$2$).
Explicitly, the formula has the form:
\be
\hat{P_i}^{+} =  b_i^{+} \sqrt{\sum_{k=0}^{\infty} \frac{(-2)^{k}}{(k+1)!}
(b_i^{+})^{k} b_i^{k}}     \qquad \hat{P_i}=(\hat{P}_i^{+})^{+} \ .
\label{AT0}
\ee
Here $\hat{P}^{+}_{i},\hat{P_{i}}$ are the creation and annihilation
operators of
a paulion on site $i$, which obey the paulionic commutation relations

$$
\hat{P}^{+}_{i}\hat{P_{i}} + \hat{P_{i}}\hat{P}^{+}_{i}=1 \qquad
(\hat{P}^{+}_{i})^{2}=\hat{P_{i}}^{2}= 0
$$
\be
[\hat{P}^{+}_{i},\hat{P_{j}}]=[\hat{P_{i}},\hat{P_{j}}]=0
\quad  \mbox{for}  \quad i \neq j
\label{a}
\ee
and $b^{+}_{i},b_{i}$ are the bosonic operators on $i$-th site.

Particles with anticommutation relations (\ref{a}) on site and the bosonic
commutation relations on different sites  widely arise in spin lattices,
magnetics, models describing
excitons in molecular crystals, defectons in quantum crystals and
many others. In the original paper~\cite{AT} the
Frenkel excitons were considered in connection with the possibility
of their Bose-condensation.
As usually in the problem of the Bose-condensation, the definition of
an auxiliary
bosonic description of the system is a central point because then
the standard  theory of a nonideal Bose-gas can be applied.

On the other hand, there are several applications of high rank truncated
oscillators. For example, it was recently shown that such operators
can be used for the second quantization of particles with Haldane
exclusion statistics~\cite{IIG}. Truncated
oscillators also find applications in nonlinear optics, semiconductors,
parasupersymmetric theories and other fields. That is why it seems to be
interesting to generalize Agranovich-Toshich description of truncated
oscillators of rank $2$ to  higher ranks and investigate the corresponding
bosonic representation.

Similar to the approach of the sigma-model with Wess-Zumino term \cite{W}
we will treat the constraint
on number of particles on site exactly. To do this we will use the
mapping of the orthogonal
sum of identical copies of a truncated oscillator space of states to the
bosonic space of
states. In such mapping the creation and annihilation
operators of truncated bosons are
represented in a form of  power series on the usual bosonic creation and
annihilation
operators. This  compels us to deal with  infinite series of different
vertices in the diagram technique. The choice of relevant contributions
in such series should be dictated as usually by feathers of the concrete
problem.

The paper is constructed as follows. In the next section we prove the
generalization of the Agranovich-Toshich formula for the case of an
arbitrary rank of  truncated oscillators. We give both variants of the
mapping -- with and without the square root (which corresponds to the formula
proposed by Chernyak~\cite{Ch} for the paulionic case; in practical use
the  latter is even more convenient). In  section 3
we describe an associated bosonic system for the case of many degrees of
freedom
and give the functional integral representation of thermodynamic quantities
of the system. The conclusion completes the letter with several remarks.

\section{ Generalization of Agranovich-Toshich representation for
truncated oscillators  }

The goal of this section is to express the creation and annihilation
operators of truncated bosons $B^+, B$ of rank $m$ with the algebra:
\be
B B^+ - B^+ B = 1 - \frac{m}{(m-1)!}(B^+)^{m-1} B^{m-1}\ , \quad (B^+)^{+}
= B \ , \quad (B^+)^{m} = B^{m} = 0
\label{alg}
\ee
in terms of the standard bosonic
creation and annihilation operators $b^+, b$. In this section
only one degree of freedom is considered but the generalization on many
degrees of freedom is straightforward and will be considered in the next
section.
First of all we will construct the number particle operator of the truncated
bosons $\hat N$ using the following operator

$$
1 +  q^{(b^+b-k)} +  q^{2(b^+b-k)} + \ldots
+  q^{(m-1)(b^+b-k)} \ ,
$$
where $q= \exp\left( i\frac{2\pi}{m}\right)$ and $m$ is the rank of
the truncated bosons. One can prove that this operator does not equal to
zero only on states $|lm+k\rangle$, where $l=0,1,2,\ldots$. So the operator
$\hat N = B^+B$ can be expressed as follows:
\ba
\hat N & = & 0\cdot\left(1 +  q^{b^+b} +  q^{2b^+b} + \ldots
+  q^{(m-1)b^+b} \right) \frac{1}{m} + \\ \nonumber
& + & 1\cdot\left(1 +  q^{(b^+b-1)} +  q^{2(b^+b-1)} + \ldots
+  q^{(m-1)(b^+b-1)} \right) \frac{1}{m} + \ldots \\ \nonumber
& + & (m-1)\left(1 +  q^{(b^+b-m+1)} +  q^{2(b^+b-m+1)} + \ldots
+  q^{(m-1)(b^+b-m+1)} \right)\frac{1}{m} \ . \nonumber
\ea
Summing terms of  same orders one can obtain the following expression
for operator $\hat{N}$:
\be
\hat N = \sum_{k=1}^{m-1}\frac{q^k}{1-q^k} q^{kb^+b} + \frac{m-1}{2} \ .
\label{N}
\ee
Now to order bosonic operators we can use the formula for the normal
ordered exponent

$$
q^{kb^+b} = \exp\left( i\frac{2\pi}{m} k b^+b\right) =
: \exp(q^k -1 )b^+b : \ ,
$$
where $:\quad :$ denotes normal ordering. Then  expression~(\ref{N})
takes the form

\be
\hat N = \sum_{l=1}^{\infty} \sum_{k=1}^{m-1}\frac{(-1)^l}{l!}
q^k (1-q^k)^{l-1} (b^+)^l b^l \ .
\label{NN}
\ee
Now let us look for the creation (annihilation) operators of truncated
bosons in the following form:

$$
B^{+} =  b^{+} \sqrt{\sum_{k=0}^{\infty} \al_k
(b^{+})^{k} b^{k}}\ ,     \qquad {B}=({B}^{+})^{+} \ .
$$
and assume that $\al_k$ are real. Then, using expression~(\ref{NN}) for
the number particle operator $\hat{N}=B^{+}B$, one can obtain the
expression for the coefficients $\al_k$:

$$
\al_l = \sum_{k=1}^{m-1}\frac{(-1)^{l+1}}{(l+1)!}
q^k (1-q^k)^{l}  \ .
$$
For $m=2$ ($q=-1$) the coefficients take the form
$$
\al_l = \frac{(-2)^l}{(l+1)!}
$$
which gives us  Agranovich-Toshich representation (\ref{AT0}) for the
creation (annihilation) operators of truncated bosons of rank 2.
So we have proved the generalization of Agranovich-Toshich formula for
truncated oscillators of higher ranks:
\be
B^{+} =  b^{+} \sqrt{\sum_{k=0}^{\infty}
\sum_{k=1}^{m-1}\frac{(-1)^{l+1}}{(l+1)!}
q^k (1-q^k)^{l}
(b^{+})^{k} b^{k}}\ ,     \qquad {B}=({B}^{+})^{+} \ , \quad
q = e^{i\frac{2\pi}{m}}
\label{AT}
\ee

Let us now "take square root" in  formula (\ref{AT}). To do this we will
follow the method proposed by Chernyak in  Ref.~\cite{Ch}. The main
point in the method is to use the projection operator on the vacuum state
of the auxiliary boson system, i.e on the vector $|0>$. To express this
projection operator ${\cal P}$ in terms of $b^{+},b$ the coherent state
representation is convenient
$$
|z>=\exp(-\frac{1}{2}\bar{z}z) \exp(zb^{+})|0> \ ,
$$
$$
<{z^{\prime}}|z>=\exp(-\frac{1}{2}\bar{z^{\prime}}z^{\prime}
-\frac{1}{2}\bar{z}z + \bar{z^{\prime}}z)  \ .
$$
{}From last relations one easily derives
$$
<{z^{\prime}}|{\cal P}|z> =
\exp(- \bar{z^{\prime}}z) <{z^{\prime}}|z>\ .
$$
On the other hand, for any $k,l$ we have
$$
<{z^{\prime}}|(b^{+})^{k}b^l|z> =
(\bar{z^{\prime}})^{k} z^{l} <{z^{\prime}}|z>\ .
$$
Then we can conclude that
$$
<{z^{\prime}}|{\cal P}|z> =
\sum_{l=0}^{\infty}\frac{(-1)^{l}}{l!}
<{z^{\prime}}|(b^{+})^{l}b^l|z> \ .
$$
It means that the projection operator has the following expression in
terms of the bosonic creation and annihilation operators:
\be
{\cal P} = \sum_{l=0}^{\infty}\frac{(-1)^{l}}{l!}
(b^{+})^{l}b^{l} \equiv :\exp(-b^{+}b): \ .
\label{P}
\ee

We now can use  formula (\ref{P}) to construct the creation and
annihilation operators $B^{+}, B$ which obey  algebra~(\ref{alg}).
Indeed, it is easy to check from the matrix form that the following
relations take place:
$$
B^{+}  =  \sum_{n=0}^{\infty} \sum_{k=0}^{m-2} (b^{+})^{mn+k+1} {\cal P}
b^{mn+k} \frac{\sqrt{k+1}}{(mn+k)!\sqrt{mn+k+1}}
$$
\be
B  =  \sum_{n=0}^{\infty} \sum_{k=0}^{m-2} (b^{+})^{mn+k} {\cal P}
b^{mn+k+1} \frac{\sqrt{k+1}}{(mn+k)!\sqrt{mn+k+1}} \ .
\label{Chernyak}
\ee
It is obvious that  relations (\ref{Chernyak}) satisfies to
algebra~(\ref{alg}). On the other hand, the operators given by relations
(\ref{AT})  satisfy the same algebra and have the same matrix form.
Hence, we can realize  formulae (\ref{Chernyak}) as the "taking of
square root" in  eq.(\ref{AT}). For the particular case $m=2$ our
formulae are reduced to the formulae originally obtained by
Chernyak~\cite{Ch} for the case of paulionic operators.

\section{Paulion-boson mapping and functional integral representation}
In this section we will describe the mapping from the system of truncated
oscillators to the auxiliary bosonic system which will be main object of the
investigation in the section. The goal is to escape the
introduction of a constraint. To do this we will embed an infinite number of
copies of the
finite dimensional space of states in the bosonic space of states and then
proceed with
the consideration of this new (auxiliary) bosonic space.

To explain this in details, let us
first of all consider one
degree of freedom (i.e a single site). Then the creation $B^{+}$ and
annihilation
$B$ operators have the following matrix form in the $m$-dimensional
Hilbert space of states ${\cal H}_B$ ($m$ is a rank of the truncated
oscillator):
$$
B^{+}=
\left(
\begin{array}{cccccc}
0 & 0 & 0 & 0 & ... & 0 \\
1 & 0 & 0 & 0 & ... & 0 \\
0 & \sqrt{2} & 0 & 0 & ... & 0 \\
0 & 0 & \sqrt{3} & 0 & ... & 0 \\
. & . & . & . & . & . \\
. & . & . & . & . & . \\
0 & 0 & 0 & 0 & \sqrt{m-1} & 0
\end{array}
\right)
$$
$$
B=
\left(
\begin{array}{cccccc}
0 & 1 & 0 & 0 & ... & 0 \\
0 & 0 & \sqrt{2} & 0 & ... & 0 \\
0 & 0 & 0 & \sqrt{3} & ... & 0 \\
. & . & . & . & . & . \\
0 & 0 & 0 & 0 & 0 & \sqrt{m-1} \\
0 & 0 & 0 & 0 & 0 & 0
\end{array}
\right)
$$
with basis $\{|0>,|1>,...,|m-1>\}$ and the obvious notations. Now we
introduce
the infinite orthogonal sum ${\cal H}_b=\oplus \sum_{n=0}^{\infty}
{\cal H}_{B,n}$ of
such $m+1$-dimensional Hilbert spaces ${\cal H}_{B,n}$ with basis
$\{\{|0>,|1>,...,|m-1>\},...,\{|nm+1>,|nm+2>,...,|nm+m-1>\},...\}$.
The extensions of
the creation and annihilation operators $B^{+},B$ in this space have the
form:
$$
\hat{B}^{+}=diag(B^{+},B^{+},...)  \qquad  \hat{B}=diag(B,B,...)  \ .
$$
There is no problem to see that all thermodynamic quantities calculated
with  operators $\hat{B^{+}},\hat{B}$ are exactly the same as ones
calculated with the original operators $B^{+},B$. Indeed, for example,
$$
<\hat{B}^{+}\hat{B}>\equiv
\frac{Sp(\hat{B}^{+}\hat{B}e^{-\beta
(E-\mu)\hat{B}^{+}\hat{B}})}{Sp(e^{-\beta(E-\mu)\hat{B}^{+}\hat{B}})}
$$
coincides with the same expressions but without hats due to the block
structure
of our operators (we should add that only the partition functions differ by
an infinite numerical constant which does not affect observable physical
quantities). The conclusion will be kept if we
start with a lattice of truncated oscillators  and then introduce hats for
the operators.

In the previous section we have found
the corresponding
expressions for the creation and annihilation operators $\hat{B}^{+},\hat{B}$
in terms of the bosonic creation and annihilation operators $b^{+},b$
acting in
the Hilbert space ${\cal H}_b$. They are given by  formula (\ref{AT})
with the square root or by  formula (\ref{Chernyak}) without it (which we
will use below).

The formulae considered above in this letter can be applied to construct
the Hamiltonian of the auxiliary bosonic system.
So if we start with the following  Hamiltonian $H_t$ of truncated
oscillators on a lattice:
\ba
H_t & = & \sum_{i} \Delta B^{+}_{i} B_{i} +
\sum_{i\neq j} M_{i,j} B^{+}_{i} B_{j} + \\ \nonumber
 & + & \sum_{i\neq j} (L_{i,j} B^{+}_{i} B_{j}^{+} + h.c.)
+ \sum_{i\neq j} J_{i,j} B^{+}_{i} B_{i} B^{+}_{j} B_{j}  \ .  \nonumber
\ea
then the corresponding Hamiltonian of the auxiliary bosons has the form:
\ba
H & = & \sum_{i} \Delta \sum_{l=0}^{\infty} a(l)
(b^{+}_{i})^{l+1} b_{i}^{l+1} +
\sum_{i\neq j} M_{i,j} b^{+}_{i} S_{ij} b_{j} + \\ \nonumber
 & + & \sum_{i\neq j} (L_{i,j} b^{+}_{i} b_{j}^{+} S_{ij} + h.c.)
+ \sum_{i\neq j} J_{i,j} \sum_{l,m=0}^{\infty}a(l)a(m)(b^{+}_{i})^{l+1}
(b^{+}_{j})^{m+1} b_{i}^{l+1} b_{i}^{m+1}  \ .
\ea
Here the following notations have been introduced:
$$
S_{i,j} =  \sum_{l,m=0}^{\infty} A(l) A(m)
(b^{+}_{i})^{l} (b^{+}_{j})^{m} b_{i}^{l} b_{j}^{m} \ ,
$$
and coefficients $A(l)$ and $a(l)$ are defined as:
$$
A(l)\equiv \sum_{k=0}^{min(m-2,l)} \sum_{n=0}^{\left[ \frac{l-k}{m} \right]}
\frac{(-1)^{l-mn-k}}{(l-mn-k)!}
\frac{\sqrt{k+1}}{(mn+k)! \sqrt{mn+k+1}}
$$
and
$$
a(l)\equiv \frac{(-1)^{l+1}}{(l+1)!}\sum_{k=1}^{m-1}
q^k (1-q^k)^{l} \ .
$$
Let us note once more that the system with  Hamiltonian
$H$ is equivalent to the original  Hamiltonian of truncated bosons $H_t$
and
does not require any additional constraint.

Using the standard procedure, we can put down the functional integral
representation of the partition function and correlators of the auxiliary
bosonic system and the original system of truncated oscillators. For
example, according to the
definition and  the formula (\ref{Chernyak}), the following relations
take place
\ba
Z & \equiv & Sp(e^{-\beta{H}}) =
\int Db^{+}(\tau)Db(\tau) e^{S} \\ \nonumber
<{B}^{+}_{i}{B}_{j}> & = & \int Db^{+}(\tau)Db(\tau)
\sum_{l,m=0}^{\infty} A(l)A(m)(b^{+}_{i}(\tau))^{l+1}
(b^{+}_{j}(\tau))^{m}b_{i}^{l}(\tau)) b^{m+1}_{j}(\tau) e^{S}/ Z \ ,\nonumber
\ea
where the action $S$ is defined by the form of the Hamiltonian $H$:
\ba
S & = & \int_{0}^{\beta} \Bigl( \sum_{i} \frac{\partial
b^{+}_{i}(\tau)}{\partial \tau} b_{i}(\tau) -
\sum_{i} \Delta \sum_{l=0}^{\infty} a(l)(b^{+}_{i}(\tau))^{l+1}
b_{i}^{l+1}(\tau) + \\ \nonumber
 & + & \sum_{i\neq j} M_{i,j} b^{+}_{i}(\tau) S_{ij}(\tau) b_{j}(\tau)
 + \sum_{i\neq j} (L_{i,j} b^{+}_{i}(\tau) b_{j}^{+}(\tau) S_{ij}(\tau) +
h.c.) + \\ \nonumber
 & + & \sum_{i\neq j} J_{i,j}
\sum_{l,m=0}^{\infty}a(l)a(m)(b^{+}_{i}(\tau))^{l+1}
(b^{+}_{j}(\tau))^{m+1} b_{i}^{l+1}(\tau) b_{j}^{m+1}(\tau) \Bigr)
d \tau \ . \nonumber
\ea
All other correlators can be obtain in the same manner and give us
the bosonic functional integral
representation which is free of  constraints and limiting procedures
(how it would be if we considered an analog of the hard-core interaction
on site and took a limit). As usually, the functional integral form allows
the simplest approach to the derivation of diagram technique rules which are
standard ones for the problems in question. It is tempting to
note that such technique is much less complicated and much more straigtforward
than spin operator one and is very natural for the consideration of problems
concerning to the Bose-condensation (Long Range Order) in the system.

\section{Conclusion}
In this letter we  considered the generalization of
Agranovich-Toshich
representation for the creation and annihilation operators of truncated
oscillators in terms of auxiliary bosons. This allowed us to formulate
the model of interacting bosons with the equivalent thermodynamic behaviour
and express various correlators of a truncated oscillator system through
series of correlators of interacting bosons.
It is important that such description is free of  constraints
or limiting procedures which take place in other approaches.

Moreover, this technique can be applied to the high spin systems or
Hubbard-like models using the
obvious transformation of truncated oscillator operators to the corresponding
spin operators or the Hubbard operators. This allows to escape the
complicated operator technique and make use only standard  one.

However, we have to note the difficulties which arise in this
framework. Indeed, we have the infinite series of types of interactions which
leads to the infinite series of the various vertices in  diagrams.
What contribution is relevant has to be determined by the concrete physical
problem
where several assumptions about the structure of a ground state and
excitations
have to be chosen. This common problem takes place for any perturbation
theory.
We think that the technique described above could be convenient in the
consideration of questions concerning the existence of Long Range Order
in a system which is  equivalent to the Bose-condensation.
Indeed, the interacting boson picture seems to be most natural for such
investigation.

\section*{Acknowledgments.}

We want to thank V.M.Agranovich, J.M.F.Gunn and M.W.Long
for the discussions of the problem.
This work was supported by the Grant of Russian Fund of Fundamental
Investigations N 95-01-00548, Euler stipend of German
Mathematical Society, INTAS-939 and by the UK EPSRC Grant GR/J35221.
We are grateful for the hospitality to the International Center for
Theoretical Physics where the work was finished.


\begin{thebibliography}{99}
\bibitem{AT}
V.M.Agranovich and B.S.Toshich, Sov.Phys. JETP {\bf 26} (1968) 104;
\bibitem{IIG}
A.V.Ilinskaya, K.N.Ilinski, G.M.F.Gunn,
{\it Fractional dimensional Hilbert space, second quantization and
dynamical interaction for particles with Halsane's exclusion statistics},
preprint TPBU-95-5;
\bibitem{Ch}
V.Ya.Chernyak, Phys.Lett. {\bf A 163} (1992) 117;
\bibitem{W}
P.Wiegmann, Phys.Rev.Lett. {\bf 60}, (1988) 821; \\
E.Fradkin, {\it Field theories of condensed matter
systems},
Addison-Wesley Publishing Company 1991.
\end{thebibliography}
\end{document}